\documentclass[a4paper,11pt]{article}
\usepackage{pos}

\usepackage[space=true]{accsupp}
\usepackage{xspace}
\usepackage{listings}
\newcommand{\code}[1]{\lstinline{#1}} 

\newcommand{\noncopynumber}[1]{%
    \BeginAccSupp{method=escape,ActualText={}}%
    #1%
    \EndAccSupp{}%
}
\newcommand{\copyablespace}{\BeginAccSupp{method=hex,unicode,ActualText=00A0}\ \EndAccSupp{}}

\definecolor{maroon}{cmyk}{0, 0.87, 0.68, 0.32}
\definecolor{halfgray}{gray}{0.55}
\definecolor{ipython_frame}{RGB}{207, 207, 207}
\definecolor{ipython_bg}{RGB}{247, 247, 247}
\definecolor{ipython_red}{RGB}{186, 33, 33}
\definecolor{ipython_green}{RGB}{0, 128, 0}
\definecolor{ipython_cyan}{RGB}{64, 128, 128}
\definecolor{ipython_purple}{RGB}{170, 34, 255}

\lstdefinelanguage{iPython}{
    morekeywords={access,and,break,class,continue,def,del,elif,else,except,exec,finally,for,from,global,if,import,in,is,lambda,not,or,pass,print,raise,return,try,while},%
    morekeywords=[2]{abs,all,any,basestring,bin,bool,bytearray,callable,chr,classmethod,cmp,compile,complex,delattr,dict,dir,divmod,enumerate,eval,execfile,file,filter,float,format,frozenset,getattr,globals,hasattr,hash,help,hex,id,input,int,isinstance,issubclass,iter,len,list,locals,long,map,max,memoryview,min,next,object,oct,open,ord,pow,property,range,raw_input,reduce,reload,repr,reversed,round,set,setattr,slice,sorted,staticmethod,str,sum,super,tuple,type,unichr,unicode,vars,xrange,zip,apply,buffer,coerce,intern},%
    sensitive=true,%
    morecomment=[l]\#,%
    morestring=[b]',%
    morestring=[b]",%
    morestring=[s]{'''}{'''},
    morestring=[s]{"""}{"""},
    morestring=[s]{r'}{'},
    morestring=[s]{r"}{"},%
    morestring=[s]{r'''}{'''},%
    morestring=[s]{r"""}{"""},%
    morestring=[s]{u'}{'},
    morestring=[s]{u"}{"},%
    morestring=[s]{u'''}{'''},%
    morestring=[s]{u"""}{"""},%
    literate=
    {á}{{\'a}}1 {é}{{\'e}}1 {í}{{\'i}}1 {ó}{{\'o}}1 {ú}{{\'u}}1
    {Á}{{\'A}}1 {É}{{\'E}}1 {Í}{{\'I}}1 {Ó}{{\'O}}1 {Ú}{{\'U}}1
    {à}{{\`a}}1 {è}{{\`e}}1 {ì}{{\`i}}1 {ò}{{\`o}}1 {ù}{{\`u}}1
    {À}{{\`A}}1 {È}{{\'E}}1 {Ì}{{\`I}}1 {Ò}{{\`O}}1 {Ù}{{\`U}}1
    {ä}{{\"a}}1 {ë}{{\"e}}1 {ï}{{\"i}}1 {ö}{{\"o}}1 {ü}{{\"u}}1
    {Ä}{{\"A}}1 {Ë}{{\"E}}1 {Ï}{{\"I}}1 {Ö}{{\"O}}1 {Ü}{{\"U}}1
    {â}{{\^a}}1 {ê}{{\^e}}1 {î}{{\^i}}1 {ô}{{\^o}}1 {û}{{\^u}}1
    {Â}{{\^A}}1 {Ê}{{\^E}}1 {Î}{{\^I}}1 {Ô}{{\^O}}1 {Û}{{\^U}}1
    {œ}{{\oe}}1 {Œ}{{\OE}}1 {æ}{{\ae}}1 {Æ}{{\AE}}1 {ß}{{\ss}}1
    {ç}{{\c c}}1 {Ç}{{\c C}}1 {ø}{{\o}}1 {å}{{\r a}}1 {Å}{{\r A}}1
    {€}{{\EUR}}1 {£}{{\pounds}}1
    {^}{{{\color{ipython_purple}\^{}}}}1
    {=}{{{\color{ipython_purple}=}}}1
    {+}{{{\color{ipython_purple}+}}}1
    {*}{{{\color{ipython_purple}$^\ast$}}}1
    {/}{{{\color{ipython_purple}/}}}1
    {?}{{{\color{ipython_purple}?}}}1
    {+=}{{{+=}}}1
    {-=}{{{-=}}}1
    {*=}{{{$^\ast$=}}}1
    {/=}{{{/=}}}1
    {\ }{{\copyablespace}}1,
    identifierstyle=\color{black}\ttfamily,
    commentstyle=\color{ipython_cyan}\ttfamily,
    stringstyle=\color{ipython_red}\ttfamily,
    keepspaces=true,
    showspaces=true,
    showstringspaces=false,
    rulecolor=\color{ipython_frame},
    frame=single,
    frameround={t}{t}{t}{t},
    framexleftmargin=6mm,
    numbers=none,
    numberstyle=\tiny\color{halfgray}\noncopynumber,
    backgroundcolor=\color{ipython_bg},
    basicstyle=\scriptsize\ttfamily,
    keywordstyle=\color{ipython_green}\ttfamily,
    aboveskip=\smallskipamount,
    belowskip=\smallskipamount
}
\definecolor{maroon}{cmyk}{0, 0.87, 0.68, 0.32}
\definecolor{halfgray}{gray}{0.55}
\definecolor{bash_frame}{RGB}{207, 207, 207}
\definecolor{bash_bg}{RGB}{247, 247, 247}
\definecolor{bash_red}{RGB}{186, 33, 33}
\definecolor{bash_green}{RGB}{0, 128, 0}
\definecolor{bash_cyan}{RGB}{64, 128, 128}
\definecolor{bash_purple}{RGB}{170, 34, 255}

\lstdefinelanguage{bash}[]{sh}%
  {morekeywords={alias,bg,bind,builtin,caller,command,compgen,compopt,%
      complete,coproc,curl,declare,disown,dirs,enable,fc,fg,help,%
      history,jobs,let,local,logout,mapfile,printf,pushd,popd,%
      readarray,select,set,suspend,shopt,source,times,type,typeset,%
      ulimit,unalias,wait},%
   otherkeywords={ [, ], [[, ]], \{, \} }%
  }%

\lstdefinelanguage{bash}{
    morekeywords={awk,break,case,cat,cd,continue,do,done,echo,elif,else,%
      env,esac,eval,exec,exit,export,expr,false,fi,for,function,getopts,%
      hash,history,if,in,kill,login,newgrp,nice,nohup,ps,pwd,read,%
      readonly,return,set,sed,shift,test,then,times,trap,true,type,%
      ulimit,umask,unset,until,wait,while},%
    morecomment=[l]\#,%
    morestring=[d]",%
    alsoletter={*"'0123456789.},%
    alsoother={\{\=\}},%
    literate={{=}{{{=}}}1},%
    literate={\$\{}{{{{\bfseries{}\$\{}}}}2,%
    otherkeywords={ [, ], \{, \} },%
    identifierstyle=\color{black}\ttfamily,
    commentstyle=\color{bash_cyan}\ttfamily,
    stringstyle=\color{bash_red}\ttfamily,
    keepspaces=true,
    showspaces=false,
    showstringspaces=false,
    rulecolor=\color{bash_frame},
    frame=single,
    frameround={t}{t}{t}{t},
    framexleftmargin=6mm,
    numbers=none,
    numberstyle=\tiny\color{halfgray},
    backgroundcolor=\color{bash_bg},
    basicstyle=\scriptsize\ttfamily,
    keywordstyle=\color{bash_green}\ttfamily,
    aboveskip=\smallskipamount,
    belowskip=\smallskipamount,
}[keywords,comments,strings]%

\newcommand{\cpp}{\texttt{C++}\xspace}
\newcommand{\EOS}{\texttt{EOS}\xspace}
\newcommand{\FlavBit}{\texttt{FlavBit}\xspace}
\newcommand{\HEPfit}{\texttt{HEPfit}\xspace}
\newcommand{\flavio}{\texttt{flavio}\xspace}
\newcommand{\Jupyter}{\texttt{Jupyter}\xspace}
\newcommand{\matplotlib}{\texttt{matplotlib}\xspace}

\newcommand{\Python}{\texttt{Python}\xspace}
\newcommand{\pypmc}{\texttt{pypmc}\xspace}

\newcommand{\SuperIso}{\texttt{SuperIso}\xspace}

\title{\EOS\ -- A Software for Flavor Physics Phenomenology}

\author*[a]{M\'eril Reboud}

\affiliation[a]{Physik Department T31, Technische Universit\"at M\"unchen, D-85748 Garching, Germany}

\emailAdd{meril.reboud@tum.de}

\abstract{I present \EOS, an open-source software dedicated to a variety of tasks in the processing of flavor physics observables.
\EOS is written in \cpp and offers both a \cpp and a \Python interface.
It is developed for three main tasks, the production of theoretical predictions for flavor physics observables;
the inference of theoretical parameters from an extensible database of likelihoods;
and the production of Monte Carlo samples of flavor processes for sensitivity studies.}

\FullConference{%
  Computational Tools for High Energy Physics and Cosmology (CompTools2021)\\
  22-26 November 2021\\
  Institut de Physique des 2 Infinis (IP2I), Lyon, France
}

\begin{document}
\maketitle
\setlength{\parindent}{0pt}

\section{Introduction}

Recent phenomenological analyses of flavor physics show a consistent pattern of tasks.
Large sets of experimental measurements are first analyzed through the prism of improved theoretical models.
New measurements are then usually suggested to further test the viability of these models, in accordance to the Standard Model or in new physics scenarios. 
These tasks mainly require
\begin{itemize}
    \item the production of publication-quality theory predictions for the experimental observables;
    \item the inference of theory parameters from an extensible database of likelihoods;
    \item and possibly the production of Monte Carlo samples for sensitivity studies.
\end{itemize}

\EOS\footnote{\url{https://github.com/eos/eos}}~\cite{EOS,EOS:paper}
has been developed since 2011~\cite{vanDyk:2012zla,EOS:repo} to perform these tasks and has already been used in about 30 peer-reviewed and published phenomenological studies~\cite{%
Bobeth:2010wg,%
Bobeth:2011gi,Bobeth:2011nj,%
Beaujean:2012uj,Bobeth:2012vn,%
Beaujean:2013soa,%
Faller:2013dwa,SentitemsuImsong:2014plu,Boer:2014kda,%
Beaujean:2015gba,Feldmann:2015xsa,Mannel:2015osa,%
Bordone:2016tex,Meinel:2016grj,Boer:2016iez,Serra:2016ivr,%
Bobeth:2017vxj,Blake:2017une,%
Boer:2018vpx,Feldmann:2018kqr,Gubernari:2018wyi,%
Boer:2019zmp,Bordone:2019vic,Blake:2019guk,Bordone:2019guc,%
Gubernari:2020eft,%
Bruggisser:2021duo,Leljak:2021vte,Bobeth:2021lya%
}.
Besides these applications in phenomenology, \EOS also is used by the collaborations of the CDF~\cite{CDF:2011tds}, the CMS~\cite{CMS:2013mkz,CMS:2015bcy} and the LHCb~\cite{LHCb:2012bin,LHCb:2013zuf,LHCb:2014auh,LHCb:2015svh,LHCb:2018jna,LHCb:2020lmf} experiments and is now part of the Belle II analysis framework~\cite{Kuhr:2018lps}.\\

\EOS is not the only openly available flavor software.
It competes, amongst others, with \flavio~\cite{Straub:2018kue}, \SuperIso~\cite{Mahmoudi:2007vz,Mahmoudi:2008tp}, \HEPfit~\cite{DeBlas:2019ehy} and \FlavBit~\cite{Workgroup:2017myk}.
The main distinctions between \EOS and these programs are:
\begin{itemize}
    \item the simultaneous inference of hadronic and new physics parameters;
    \item the modularity of hadronic matrix elements, i.e., the possibility to select from various hadronic models and parametrizations at run time;
    \item the production of pseudo events for sensitivity studies; and
    \item the implementation of QCD sum rules for the prediction of hadronic matrix elements.\\
\end{itemize}

\EOS can be installed using \Python package installer:
\begin{lstlisting}[language=bash]
python3 -m pip install --user eoshep
\end{lstlisting}
The \EOS \Python module can then be accessed, e.g. within a \Jupyter notebook, using
\begin{lstlisting}[language=ipython]
import eos
\end{lstlisting}
\EOS documentation \cite{EOS:doc, EOS:paper} includes basic tutorials, detailed examples for advanced use, and automatically updated lists of observables, parameters and constraints.

\section{Usage and examples}

\subsection{Predictions and Uncertainties}

Observables are one of the main classes in \EOS.
They are usually defined for several theoretical models, modifiable at run time via a set of options.
The numerical evaluation of observables requires a kinematic specification and a set of values for all the parameters.
\begin{lstlisting}[language=iPython]
eos.Observable.make('B->Dlnu::BR', eos.Parameters(),
                                   eos.Kinematics({'q2_min': 0.01, 'q2_max': 11.62}),
                                   eos.Options({'l': 'mu', 'model': 'SM'})).evaluate()
\end{lstlisting}
Here, the integrated branching ratio of $B\to D\mu\nu_\mu$ is evaluated between $0.01$ and $11.62~\mathrm{GeV}^2$ in the Standard Model with \EOS default parameters.
To ensure a fast numerical evaluation of observables, \EOS uses multiple threads and reuses objects that are shared between multiple observables.
An updated list of built-in observables and parameters can be found in the online documentation \cite{EOS:doc}.

\begin{figure}[t]
    \centering
    \includegraphics[width=0.48\linewidth]{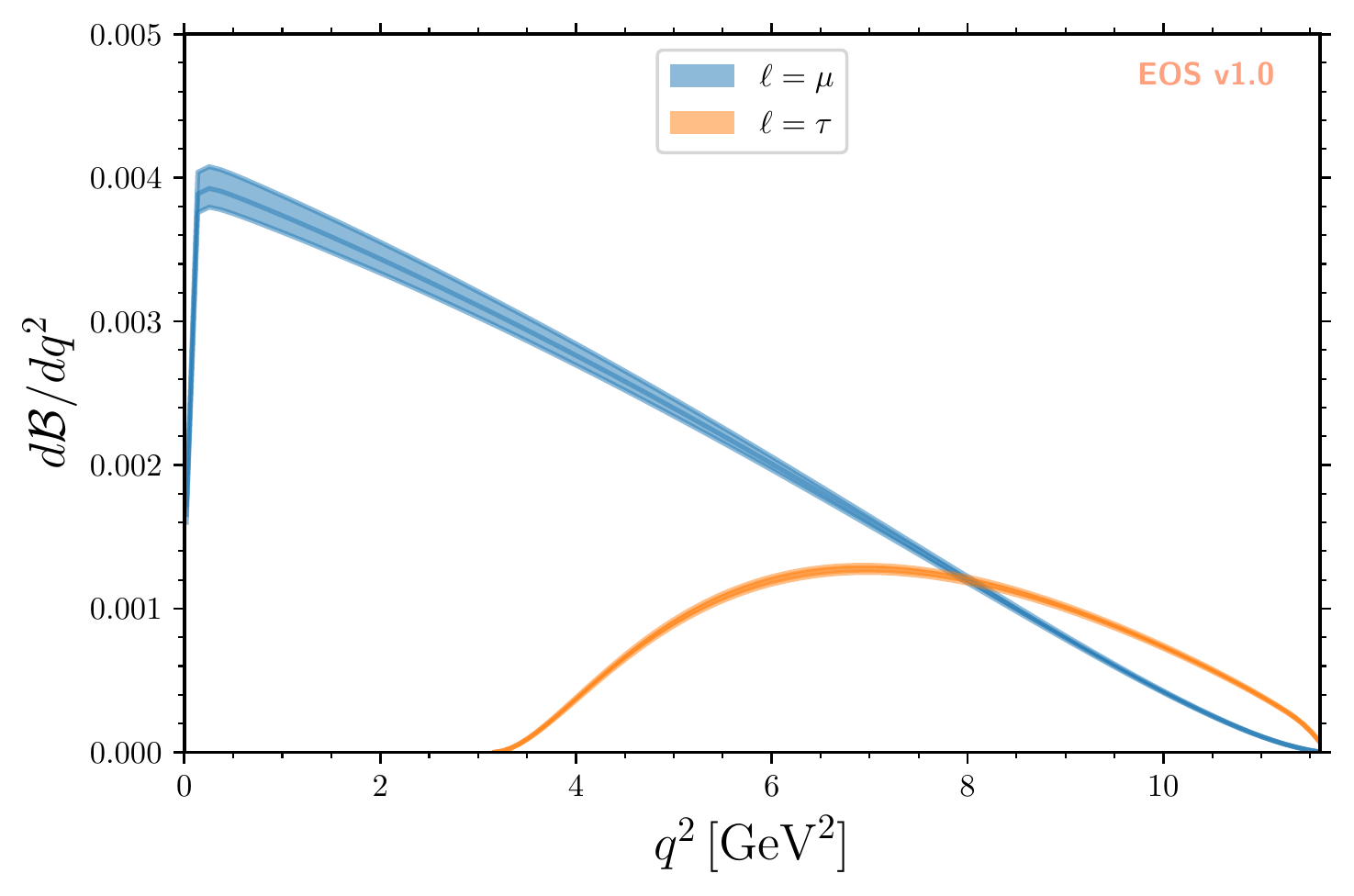}
    \caption{Differential branching ratio of $B\to D\ell\nu_\ell$ for different leptons.
    The uncertainty bands contain 68\% of the samples obtained by varying the parametrization of the hadronic form factors.}
    \label{fig:prediction}
\end{figure}

The visualization of observables can also be performed via a versatile \matplotlib-based~\cite{matplotlib} plotting framework.
Evaluating the differential observable \code{B->Dlnu::dBR/dq2} for several values of $q^2$ yields, for example, the middle solid lines of Figure~\ref{fig:prediction}.\\

\EOS bases the estimation of theory uncertainties on Monte Carlo techniques and relies on the external \pypmc library \cite{pypmc}.
The sampling of the probability density functions is performed using adaptive Metropolis-Hastings~\cite{doi:10.1063/1.1699114,10.1093/biomet/57.1.97,10.2307/3318737} and Population Monte Carlo (PMC)~\cite{2010MNRAS.405.2381K,2013arXiv1304.7808B} sampling.
Once the user has provided the set of parameters to be varied and the experimental or theoretical likelihoods to constrain them, samples can be drawn from the joint posterior to predict uncertainties for the observables.
Pursuing with the $B\to D\ell\nu$ example, the main source of uncertainty is due to the hadronic form factors that describe the $B\to D$ transition.
Using the parametrization of Ref.~\cite{Straub:2015ica} and independent constraints obtained from lattice QCD simulations by the HPQCD \cite{Na:2015kha} and FNAL/MILC \cite{Lattice:2015rga}, we obtain the uncertainty band presented in Figure~\ref{fig:prediction}.

A list of built-in constraints can be found online \cite{EOS:doc}; new constraints can also be added via the \code{manual\_constraints} method, as also described in the documentation.\\

\subsection{Parameter Inference}

Parameters inference is theoretically equivalent to uncertainty estimation, and both are treated in the same way in \EOS.
The parameters of interest are added to the list of varied parameters (which become nuisance parameters) and the experimental measurements from which parameters are to be inferred are added to the list of likelihoods.
The posterior distribution of the parameters can again be explored using Monte Carlo techniques.

In the case of multimodal distributions, a single Markov chain is usually insufficient to explore the entire posterior distribution.
\EOS therefore implements PMC sampling, where an initial proposal distribution (obtained for example by running multiple Markov chains) is adjusted stepwise to match the posterior distribution.
This allows the user to produce high quality, statistically uncorrelated samples from the posterior distribution.

For example, Belle measurements of $B\to D\ell\nu$ differential branching ratios can be used to extract the Cabibbo-Kobayashi-Maskawa matrix element $|V_{cb}|$ \cite{Belle:2015pkj}.
The posterior samples are genuine \Python array and can be analyzed using \EOS plotting framework as presented in Figure~\ref{fig:prediction}, left panel.
The uncertainties obtained on $|V_{cb}|$ and on the observables include both the experimental uncertainties due to branching ratio measurements and the theoretical uncertainties due to the hadronic form factors.

\begin{figure}[t]
    \centering
    \includegraphics[width=0.48\linewidth]{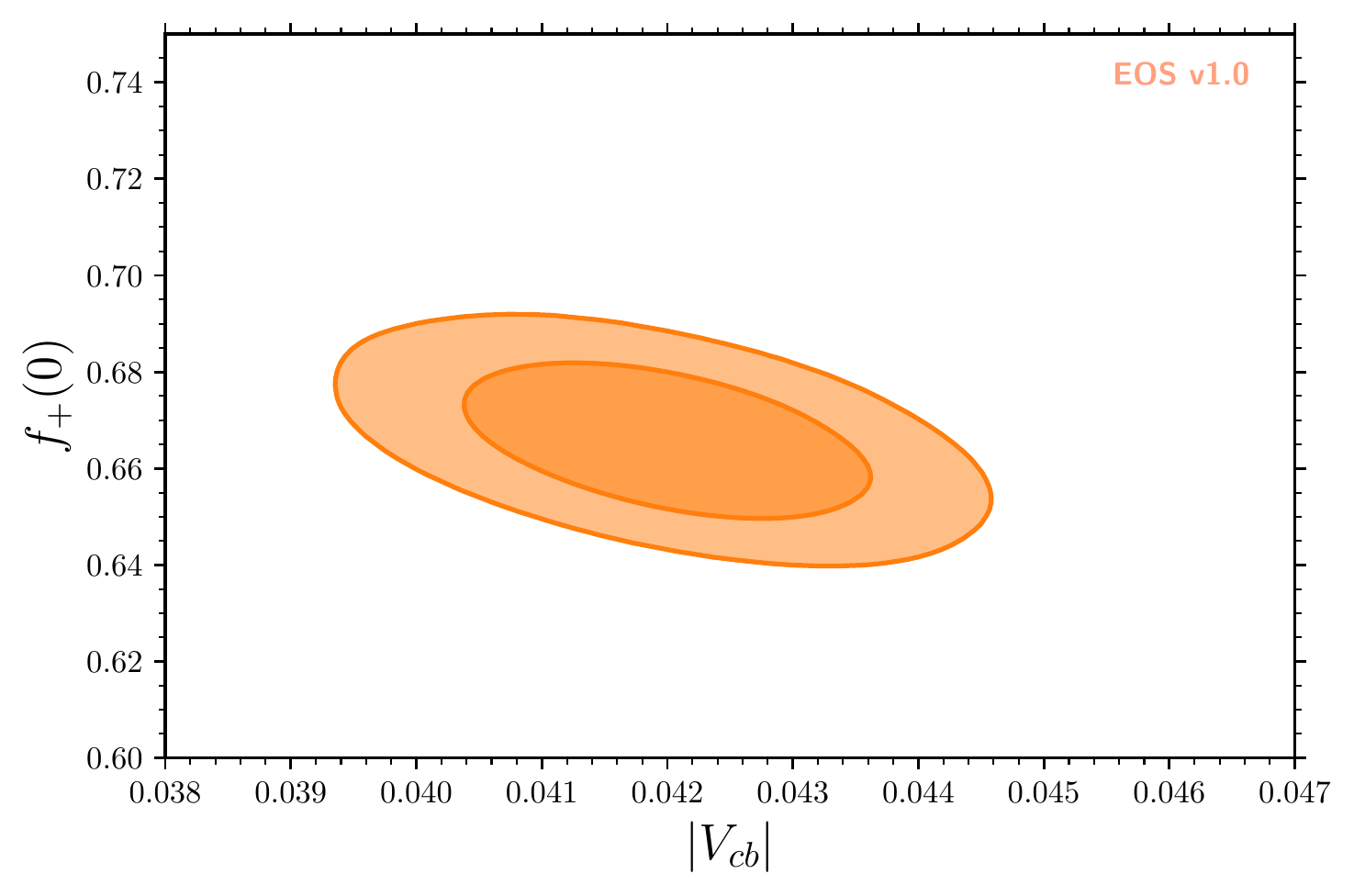}
    \includegraphics[width=0.48\linewidth]{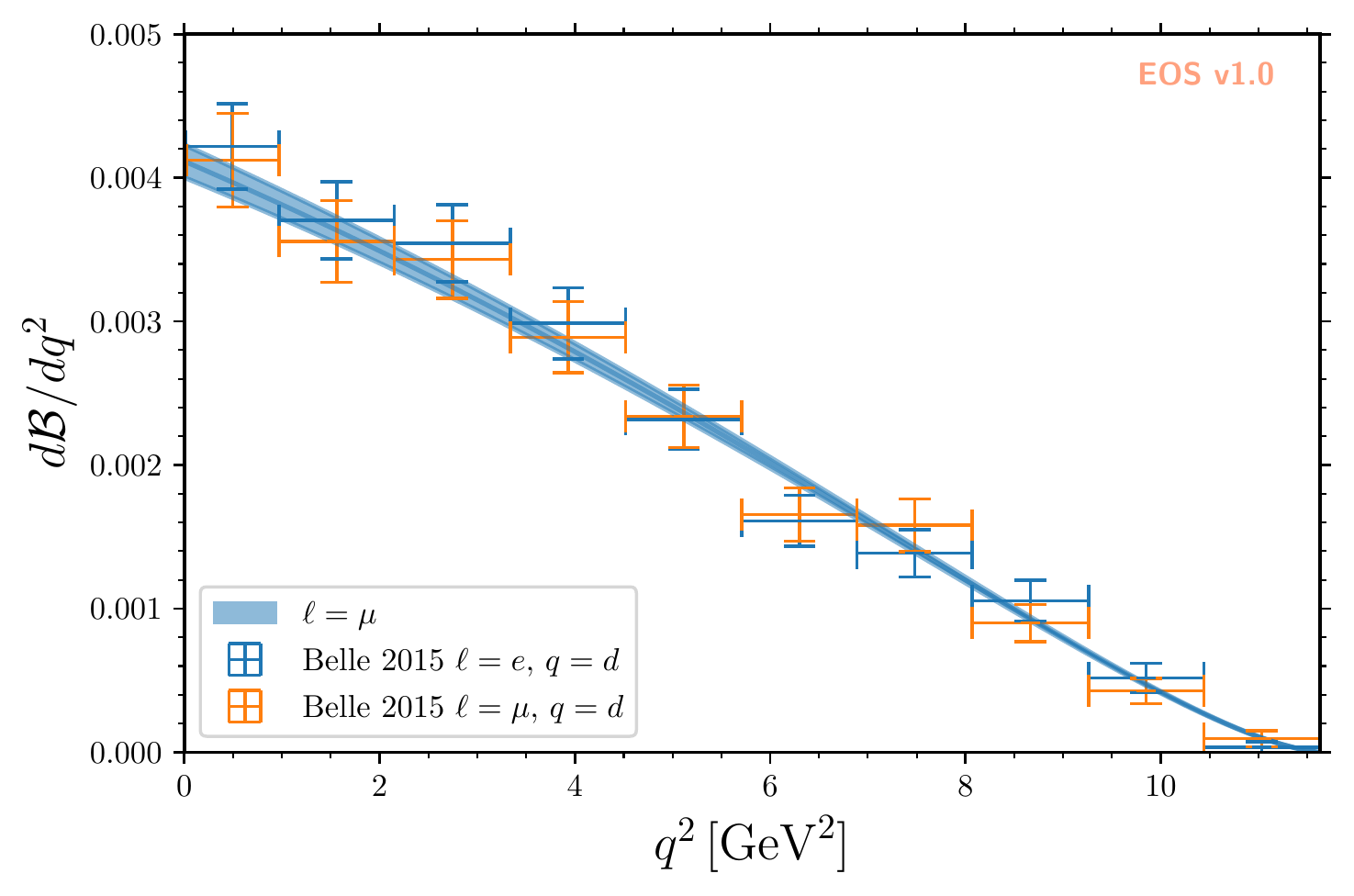}
    \caption{Inference of $|V_{cb}|$ from experimental measurements of $B\to D\ell\nu$: (left) 2D-marginal joint posterior of $|V_{cb}|$ and $f^{\bar{B}\to D}_+(0)$ ($68\%$ and $95\%$ probability contours) and (right) juxtaposition of the bin-averaged measurements of $B\to D\ell\nu$ and the 68\% uncertainty band estimated by sampling the posterior distribution.}
    \label{fig:inference}
\end{figure}

\subsection{Simulation of Pseudo-events}

\begin{figure}[t]
    \centering
    \includegraphics[width=0.48\linewidth]{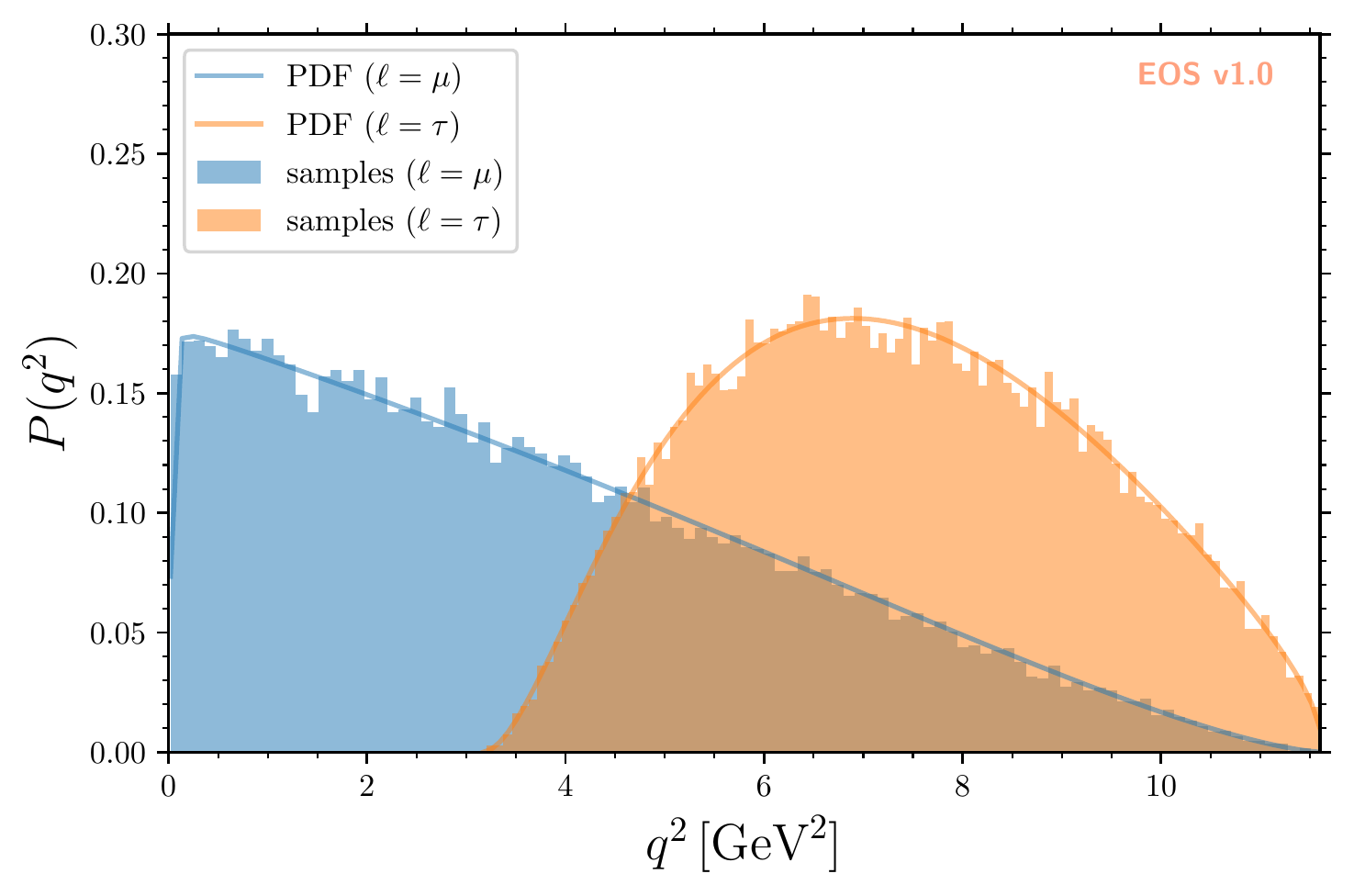}
    \caption{Distribution of $B\to D\ell\nu_\ell$ events for $\ell=\mu, \tau$, as implemented in \EOS (solid lines) and as obtained from Markov Chain Monte Carlo importance sampling (histograms).}
    \label{fig:simulation}
\end{figure}

Once a model is defined, it is often useful to investigate the experimental sensitivity to new observables that show, for example, a reduced theoretical uncertainty.
\EOS therefore contains built-in probability density functions (PDF) from which pseudo-events can be simulated.
To conclude the $B\to D\ell\nu$ example, we generate samples from the one-dimensional PDF that describes the $q^2$-differential decay distribution for $\ell=\mu$ and $\ell=\tau$.
The samples are shown in Figure~\ref{fig:simulation} overlaid with the implemented PDF for which excellent agreement is found.

\section{Conclusion}

\EOS is a multipurpose flavor physics software.
Its large and constantly growing number of built-in observables, parameters and constraints\footnote{The updated lists can be found on \EOS website \cite{EOS:doc}} allows a very wide spectrum of studies.
These studies range from the inference of theory parameters from experimental measurements to sensitivity studies of new observables in new physics scenarios.

\EOS developers welcome new contributors, feedback, questions and wishes on \url{https://github.com/eos/eos}.

\bibliography{%
bibliography/cs.bib,%
bibliography/physics.bib,%
bibliography/statistics.bib,%
bibliography/lhcb.bib%
}
\bibliographystyle{bibliography/JHEP.bst}

\end{document}